\documentclass[aps,pre,twocolumn]{revtex4}
\usepackage{graphicx}
\usepackage{epsfig}
\usepackage{amsmath}
\usepackage{graphics}
\usepackage{epsfig}
\usepackage{tikz}
\usetikzlibrary{decorations.pathmorphing}

\newcommand{\bra}[1]{\ensuremath{\langle{#1}|\,}}
\newcommand{\ket}[1]{\ensuremath{\,|{#1}\rangle}}

\begin{document}

\title{Waiting time distribution for electron transport in  a molecular junction with  electron-vibration interaction}
\author{Daniel S. Kosov}
\address{College of Science and Engineering, James Cook University, Townsville, QLD, 4811, Australia }


\begin{abstract}
On the elementary level, electronic current consists of individual electron tunnelling events that are separated by random time intervals. The waiting time distribution is a probability to observe the electron transfer in the detector electrode at time $t+\tau$ given that an electron was detected in the same electrode at earlier time $t$. We study waiting time distribution for quantum transport in a vibrating molecular junction. By treating the electron-vibration interaction exactly and molecule-electrode coupling perturbatively, we obtain  master equation and compute the distribution of waiting times for electron transport. The details of waiting time distributions are used to elucidate microscopic mechanism of electron transport and the role of electron-vibration interactions. We find that as nonequilibrium develops in molecular junction, the skewness and dispersion of the waiting time distribution experience stepwise drops with the increase of the electric current.  These steps are associated with the excitations of vibrational states by tunnelling electrons. In the strong electron-vibration coupling regime, the dispersion decrease dominates over all other changes in the waiting time distribution as the molecular junction departs far away from the equilibrium.
\end{abstract}

\maketitle
\section{Intruduction}
The basic building block for molecular electronic devices is a single molecular junction - a molecule attached to two metal electrodes held at different electronic chemical potentials.
One of the most distinct features  of molecular junctions in comparison to other nanoscale electronic devices is the absence of the structural rigidity and, as a result, the strong interplay between electronic and nuclear dynamics \cite{moletronics,galperin07}. Electron-vibration interaction leads to a variety of interesting transport phenomena such as negative differential resistance \cite{PhysRevB.83.115414,kuznetsov-ndr,galperin05,zazunov06,solvent12}, Frank-Condon blockade \cite{PhysRevB.74.205438,PhysRevB.73.155306},  current induced chemical reactions\cite{dzhioev11,PhysRevB.86.195419,catalysis12}, cooling of nuclear motion by electric current \cite{galperin09,Ioffe:2008aa,PhysRevB.83.115414}

 Recently, there have been a significant experimental and theoretical interest in studying electron molecular electron  transport properties which go beyond average electric current.
Investigation of  noise, full counting statistics, and fluctuation relations have been recently reported
 \cite{PhysRevB.90.075409,PhysRevB.92.245418,PhysRevB.84.205450,PhysRevB.87.115407,PhysRevB.91.235413,avriller09,thoss14,segal15}.
Electrical current fluctuations in molecules is no longer just a theoretical concept, they have become the important experimental method to characterize the physical mechanisms of electron transport in molecular junctions \cite{PhysRevLett.100.196804,doi:10.1021/nl201327c,doi:10.1021/nl060116e,Tsutsui:2010aa}.

 Electron transport through a molecular junction is unavoidably stochastic due to quantum nature of the process.  
Electrons are transferred from source to drain electrode across the molecular bridge one by one at random but specific times. What are the delay times  between these electron tunnelling events and what are the distributions of these times? How does coupling of electronic and nuclear motion manifest itself in this distribution? These are the questions which interest us in this paper.
The sequential stochastic processes, in general, and quantum electron transport, in particular, can be naturally  understood and characterised  in terms of  waiting time distribution (WTD) \cite{vanKampen}. WTD is a  conditional  probability distribution 
 that  we observe  the electron transfer in the detector electrode (it does not matter drain or source electrodes, in the steady state they measure the same statistics)  at time $t +\tau$ given that an electron was detected in the same electrode   at  time $t$.
 WTD is the complementary  and much more intuitive physical quantity in comparison with very popular full counting statistics in quantum transport.  WTD has recently gained significant popularity in the study current fluctuations in nanoscale and mesoscale systems 
 \cite{brandes08,buttiker12,flindt13,sothmann14,flindt14,flindt15,wtd-transient,PhysRevB.92.125435,harbola15,rudge16a, rudge16b}. 

The paper is organised as follows. Section II contains the derivation of master equation for electron transport through a molecular junction described by Anderson- Holstein model. In section III, we derive the main equations for WTD and waiting time cumulant-generating function. Section IV describes the results of numerical calculation. Section IV summarises the main results of the paper.

We use natural units in equations throughout the paper: $\hbar=k_{B}=e=1$.

\section{Master equation in the polaronic regime}

The molecule is described by spinless Anderson-Holstein model in which a single electronic level is coupled to a localised vibrational mode. Electrons can tunnel between the molecule and source (S) and  drain (D) electrodes. 
The corresponding Hamiltonian is
\begin{equation}
H= H_{\text{molecule}} + H_{\text{electrodes}} + H_T.
\end{equation}
The molecule  is described by the following Hamiltonian:
\begin{equation}
H_{\text{molecule}}= \epsilon_0 a^\dag a  + \lambda (b^\dag + b) a^\dag a +  \omega b^\dag b,
\end{equation}
where $\epsilon_0$ molecular orbital energy,  $\omega$ is molecular vibration energy, and $\lambda$ is the strength of the electron-vibration coupling. $a^\dagger (a) $  creates (annihilates) an electron on molecular orbital, and $b^+ (b)$ is bosonic creation (annihilation) operator for the molecular vibration. The electronic spin does not play any physical role in the discussed processes and will not be included explicitly into the equations.
Electrodes consist of noninteracting electrons:
\begin{eqnarray}
H_{\text{electrodes}}=  \sum_{k,\alpha=S,D} \epsilon_{k\alpha} a^\dag_{k \alpha} a_{k \alpha},
\end{eqnarray}
where $a^\dagger_{k\alpha}$   creates  an electron in the single-particle state $k$ of the source(drain) electrode  $\alpha=S(D)$ and $a_{k\alpha}$ is the corresponding electron  annihilation operator. The bias voltage applied to the junction is imposed by shifting symmetrically the chemical potentials of the electrodes $V_{sd} = \mu_S -\mu_D$. The electron tunnelling is described by 
\begin{eqnarray}
H_T=  \sum_{k,  \alpha=S,D } t_\alpha (  a^\dag_{k \alpha} a  + h.c),
\end{eqnarray}
where $t_\alpha$ is the tunnelling amplitudes.

To eliminate electron-vibration coupling from $H_{\text{molecule}}$ we perform Lang-Firsov unitary rotation of molecular operators \cite{lang_firsov1963} 
\begin{equation}
a= \tilde a e^{\nu(\tilde b^\dag -\tilde b)}, \;\;\;\; b = \tilde b + \nu \tilde a^\dag \tilde a,
\end{equation}
where $\tilde a^\dag (\tilde a )$ and $\tilde b^\dag (\tilde b)$ are  transformed  creation (annihilation) operators for molecular electron and vibration.
The Lang-Firsov transformation is unitary so that it preserves commutation or anticommutation relations between the operators.
The molecular Hamiltonian in the transformed operator basis becomes
\begin{equation}
H_{\text{molecule}} = \epsilon \tilde a^\dag \tilde a + \omega \tilde b^\dag \tilde b.
\end{equation}
Here renormalised molecular orbital energy $\epsilon$   includes polaron shift
$\epsilon = \epsilon_0 - \lambda^2/ \omega $.
The Hamiltonian for the electrodes  is not affected and the tunnelling interaction becomes
\begin{eqnarray}
H_T=   \sum_{k \alpha  } t_\alpha ( e^{-\frac{\lambda}{\omega} (\tilde b^\dag -\tilde b)} a^\dag_{k\alpha} \widetilde a  + h.c) 
\end{eqnarray}
After Lang-Firsov transformation the eigenvectors  and eigenenergies of molecular Hamiltonian are easily computed
and standard theoretical methods are applied to derive the master equation \cite{PhysRevB.69.245302}:
\begin{eqnarray}
\dot P_{0q}(t) &=& \sum_{q'} \Gamma_{1q',0q} P_{1q'} (t)  -  \Gamma_{0q,1q'} P_{0q}(t),
\label{me1}
\\
\dot P_{1q}(t)&=& \sum_{q'} \Gamma_{0q',1q} P_{0q'}(t)  -  \Gamma_{1q,0q'} P_{1q}(t),
\label{me2}
\end{eqnarray}
where  $P_{nq}(t)$ is the probability that the molecule is occupied by $n$ electrons and populated by $q$ vibration at time $t$. The  rates for the transition from state occupied by one electron and $q$ vibrations to the electronically unoccupied state with $q'$ vibrations  by the electron transfer from the molecule to $\alpha=S,D$  electrode is
$ \Gamma^\alpha_{1q,0q'} $ and 
the rate for the transition when electron is transferred from $\alpha$  electrode into the originally empty molecules simultaneously changing the vibrational state from $q$ to $q'$ is
$\Gamma^\alpha_{0q,1q'}$.
The total probability is normalised 
\[
\sum_{q}  P_{0q}(t) + P_{1q}(t) =1.
\]

The transition rates rates are  computed using the Fermi-golden rule \cite{PhysRevB.69.245302}.
They are  
\begin{equation}
\Gamma^\alpha_{1q,0q'} = \gamma^\alpha_{q'q} \left(1-f_\alpha[\epsilon-\omega (q'-q)] \right).
\end{equation}
and
\begin{equation} 
\Gamma^\alpha_{0q,1q'} = \gamma^\alpha_{q'q} f_\alpha[\epsilon-\omega (q'-q)].
\end{equation}
The rates depends on the the occupation of electrodes  given by Fermi-Dirac numbers
\begin{equation}
f_\alpha(E) = \frac{1}{1+e^{(E-\mu_\alpha)/T}},
\end{equation}
where $T$ is the temperature and $\mu_\alpha$ is the chemical potential of the electrode $\alpha$.  The transition rates are proportional to 
\begin{equation}
\gamma^\alpha_{q'q}= 2 \pi t_\alpha ^2 |X_{q'q}|^2 \rho_\alpha,
\end{equation}
where $\rho_\alpha$ is density of states in the electrode $\alpha$ taken at energy $\epsilon$ and
\begin{equation}
X_{qq'}= \bra{q} e^{-\lambda/\omega (b^\dag  -b)} \ket{q'} .
\end{equation}
is the Frank-Condon factor.

We introduce vector of probabilities, which is ordered such that the electronic populations enter in pairs for each vibrational  sub-bands:
\begin{eqnarray}
\mathbf P(t) =
\begin{bmatrix}
P_{00} (t)\\
P_{10}(t)\\
{P_{01}(t)}\\
{P_{11}(t)}\\
\vdots
\\
{P_{0N}(t)}\\
{P_{1N}(t)}\\
\end{bmatrix} ,
\end{eqnarray}
where $N$ is the total number of vibrational sub-bands included into the calculations.
We also define the identity  vector of length $2N$:
\begin{eqnarray}
\mathbf I =
\begin{bmatrix}
1\\
1\\
1\\
1\\
\vdots
\\
1\\
1\\
\end{bmatrix} .
\end{eqnarray}
The normalisation of probability is given by the scalar product between $\mathbf I $ and $\mathbf P $ vectors $ ( \mathbf I,  \mathbf P(t))$, which is  defined in a standard mathematical way as
\begin{equation}
 ( \mathbf I,  \mathbf P(t))= \sum_{q=0}^N P_{0q}(t) + P_{1q}(t)  =1.
\end{equation}

\section{Quantum jumps operators for electron tunnelling and waiting time distributions}

Let us re-write the master equation (\ref{me1},\ref{me2}) in the matrix form
\begin{equation}
\dot {\mathbf P}(t) = {\cal L} \mathbf P(t),
\end{equation}
where $\cal L$ is the the total Liouvillian operator, which can be explicitly identified using eqs. (\ref{me1},\ref{me2}). We also define 4 different quantum jumps operators  $J_+^S$, $J_-^S$, $J_+^D$, and $J_-^D$ for the transitions involving changes of molecular electronic population. The quantum jump operators are $2N\times 2N$   matrices which are defined through their actions on the probability vector:
\begin{equation}
(J^\alpha_{+} \mathbf P(t) )_{mq} =  \delta_{m1} \sum_{q'} \Gamma_{0q',1q} P_{0q'}(t),
\end{equation}
\begin{equation}
(J^\alpha_{-} \mathbf P(t) )_{mq} =  \delta_{m0} \sum_{q'} \Gamma^{\alpha}_{1q',0q} P_{1q'} (t). 
\end{equation}
The jump operator $J_{+}^{\alpha}$  transforms the system from the electronically empty  state to 
the singly occupied state by tunnelling of an electron from the $\alpha$ electrode into the molecule.
The jump operator $J_{-}^{\alpha}$ describes the reverse process: it  transforms the molecule from being occupied by one electron to being 
empty by transferring one electron from the molecule  to the $\alpha$  electrode.

After these preliminary definitions and rearrangement of the rate equation, we are ready to derive the expression for WTD for each quantum jump operator.
We assume that 
 the system has evolved to the nonequilibrium steady state. That means it is described by the steady state density matrix, which is the null vector of the full Liouvillian --
\begin{equation}
  {\cal L} \; \mathbf P =0.
\end{equation}
Let us begin to monitor time delays between sequential quantum tunnelling in the nonequilibrium steady state. 
First, we define  {\it WTD as  the conditional  probability distribution 
 that  we observe electron tunnelling  $J_{\pm}^{\alpha}$  at time $t +\tau$ given that the molecule undergoes the same quantum jump $J_{\pm}^{\alpha}$  at earlier time $t$ }
\begin{eqnarray}
w^\alpha_\pm(\tau) = ( \mathbf I, \; J^\alpha_\pm \; e^{ ({\cal L} -J^\alpha_\pm ) \tau} \; J^\alpha_\pm  \; \mathbf P ).
\label{wtd1}
\end{eqnarray}
WTD does not depend on the reference time $t$ in the steady state regime.
Reading this equation from right to left elucidates its physical meaning. The system is prepared in state described by steady state probability vector $\mathbf P $, it undergoes quantum jump $J^\alpha_\pm$  at some arbitrary time, then the system evolves without experiencing any of the monitored quantum jumps  for time $\tau$ (this dynamics is generated by nonunitary evolution operator $e^{ ({\cal L} -J^\alpha_\pm ) \tau}$) and finally it undergoes the quantum jump $J^\alpha_\pm$. In the end, by computing the  scalar product with vector $\mathbf I$, the resulting probability vector is summed over all electronic and vibrational states in order to give the total probability distribution for this event.
Since we propagate the system with $e^{ ({\cal L} -J^\alpha_\pm ) \tau}$,  there were no other electron transfer events  {\it  of the same type $J^\alpha_\pm$} between time $t$ and $t+\tau$.

WTD (\ref{wtd1}) is not normalised yet. We assume that over all time $0 \le \tau \le +\infty$ the probability for a quantum jump to occur is unity. Integrating over the waiting time yields
\begin{eqnarray}
\int_0^{\infty} d \tau\; w^\alpha_\pm(\tau) = ( \mathbf I, \; J^\alpha_\pm \; (J^\alpha_\pm - {\cal L}  )^{-1} \; J^\alpha_\pm  \; \mathbf P ) =
\nonumber
\\
=   ( \mathbf I, \; (J^\alpha_\pm -{\cal L} + {\cal L})  \; (J^\alpha_\pm - {\cal L}  )^{-1} \; J^\alpha_\pm  \; \mathbf P ) =  (\mathbf I, \;  J^\alpha_\pm \;  \mathbf P).
\nonumber
 \end{eqnarray}
 Here we take into account (easy to prove from the conservation of the probability)  property of the Liouvillian that $(\mathbf I, \; {\cal L} \mathbf X)=0$ for an arbitrary vector $\mathbf X$ \cite{rudge16a}.
 The normalized WTD becomes:
\begin{equation}
w^\alpha_\pm(\tau) = \frac{  (\mathbf I, \; J^\alpha_\pm \; e^{ ({\cal L} -J^\alpha_\pm ) \tau} \; J^\alpha_\pm \;  \mathbf P) }{  (\mathbf I, \;  J^\alpha_\pm \;  \mathbf P)}. 
\label{wtd2}
\end{equation}
We have 4 different  WTDs associated with each 4 quantum jump operators for electron tunnelling $J_+^S$, $J_-^S$, $J_+^D$, and $J_-^D$.

To compute higher-order expectation values and analyse the fluctuations, it is convenient to introduce the cumulant-generating function for the WTD  as
\begin{eqnarray}
 K^\alpha_\pm(x)= \int_0^\infty d \tau e^{x \tau} w^\alpha_\pm(\tau).
 \end{eqnarray}
 This expression can be further simplified and brought to the form suitable for numerical calculations:
 \begin{eqnarray}
\nonumber 
&& K^\alpha_\pm(x)= \frac{  (\mathbf I, \; J^\alpha_\pm \;  \int_0^\infty  d \tau e^{ ({\cal L} -J^\alpha_\pm+x  ) \tau} \; J^\alpha_\pm \;  \mathbf P) }{  (\mathbf I, \;  J^\alpha_\pm \;  \mathbf P)} 
\nonumber
\\
&& = - \frac{  (\mathbf I, \; J^\alpha_\pm \;  ({\cal L} -J^\alpha_\pm+x  )^{-1}  \; J^\alpha_\pm \;  \mathbf P) }{  (\mathbf I, \;  J^\alpha_\pm \;  \mathbf P)}. 
\label{cum}
\end{eqnarray}
We obtain all possible higher order comulants simply by direct differentiation of $K^\alpha_\pm(x)$ with respect to $x$.

\section{Results}

\begin{figure}[t!]
\begin{center}
\includegraphics[width=\columnwidth]{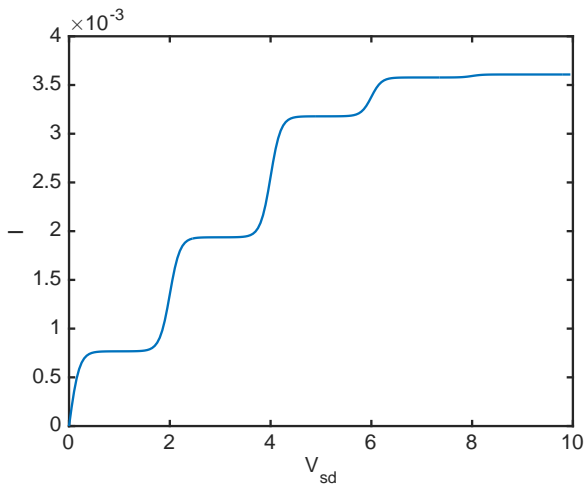}
\end{center}
	\caption{Current $I$ as a function of applied voltage  $V_{sd}$. Parameters used in calculations (all energy values are given in units of $\omega$): $\gamma_S = \gamma_D =0.01$, $ T =0.05$, 
$\epsilon =0$, $\lambda=1 $. Unit for for electric current is $\omega$ (or if we put $\hbar$ and $e$ back, it is  $ e \omega$) and  values of voltage bias $V_{sd}$ are given in $\omega$ (or $\hbar \omega/e$).}	
\label{IV}
\end{figure}

We first compute electric current as a function of the applied voltage bias $V_{sd}= \mu_S- \mu_D$.  Fig.\ref{IV} shows the current-voltage characteristics. It has been studied intensively in numerous works before \cite{galperin07,PhysRevB.83.115414,moletronics} and we show it here simply to serve as a reference - the characteristics steps in the current-voltage characteristic  will be shortly connected to the  behaviour of WTDs. The steps  in the current is related to the resonant excitations of the vibration states by electric current which occur when the voltage passes through an integer multiple  of the vibration energy.  These steps are smoothed due to the temperature effects.

\begin{figure}[t!]
\begin{center}
\includegraphics[width=\columnwidth]{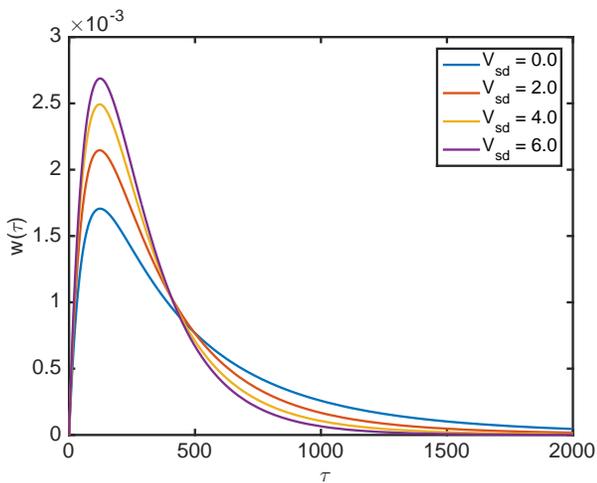}
\end{center}
	\caption{WTD between the detection of the transferred electron in the drain electrode $w_-^D(\tau)$. Parameters used in calculations (all energy values are given in units of $\omega$): $\gamma_S = \gamma_D =0.01$, $ T =0.05$, 
$\epsilon =0$, $\lambda=1 $. Time $\tau$ is measured in  $1/\omega$.}	
\label{wtd1-fig}
\end{figure}

WTD between the detections 
of electrons transferred from the molecule to the drain electrode
 is shown in Fig.\ref{wtd1-fig}. We plot $w_-^D$  as a function of $\tau$ for different values of the applied voltage. It is computed with the use of (\ref{wtd2}). 
For molecular junction symmetrically coupled to the source and drain electrodes $\gamma_S=\gamma_D$ this quantity is exactly the same as the WTD for electron transfer from the source electrode to the molecule, $w_+^S$ .
 The interesting feature of this distribution is that  it becomes less spread as the voltage increases whereas the mode of the distribution, that is the value of waiting time between electron transfer events that appears most often in the electron transport - it corresponds to the peak of the distribution, remains more or less constant. 
 
\begin{figure}[t!]
\begin{center}
\includegraphics[width=\columnwidth]{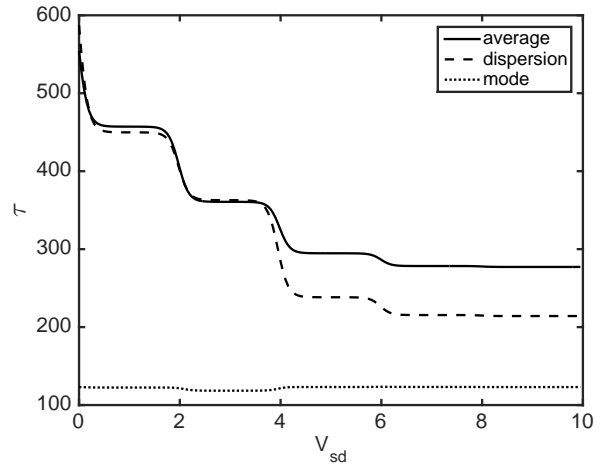}
\end{center}

\caption{Average waiting time, dispersion and  mode time   between the detection of an  electron tunnelling from  the molecule to  the drain electrode. Parameters used in calculations (all energy values are given in units of $\omega$): $\gamma_S = \gamma_D =0.01$, $ T=0.05$, 
$\epsilon =0$, $\lambda=1 $. Time $\tau$ is measured in  $1/\omega$ and  values of voltage bias $V_{sd}$ are given in $\omega$.}
	
\label{avt1}
\end{figure}

 Fig.\ref{avt1} illustrates this observation more directly by 
showing  the average waiting time between electron detection in the drain electrode, dispersion $\sqrt{\langle \tau^2 \rangle - \langle \tau \rangle^2} $  and mode time for  WTD $w_-^D$ as functions of applied voltage. The mode time is almost not affected by the voltage and  remains approximately the same as the equilibrium value.  Therefore, if we monitor electric current spikes that most often observed time delays  do not depend on voltage at all.  The average waiting time (the difference between mode time and average time can be associated with skewness of the distribution) decreases as the voltage grows and shows the characteristic step behaviour related to   excitation of the vibrational states by the tunnelling electrons. We see  similar but much more pronounced behaviour for the dispersion of the distribution. If population of the vibrational states does not change, the WTD remains the same even if the voltage increased. When the number of excited vibrations rises by one quanta, the WTD is squeezed in step-like fashion (both skewness and dispersion shrinks abruptly when the vibrational state is excited). Therefore, the microscopic mechanism of current growth is the reduction of the long tail of slow electrons by making the WTD narrower around its mode value.

\begin{figure}[t!]
\begin{center}
\includegraphics[width=\columnwidth]{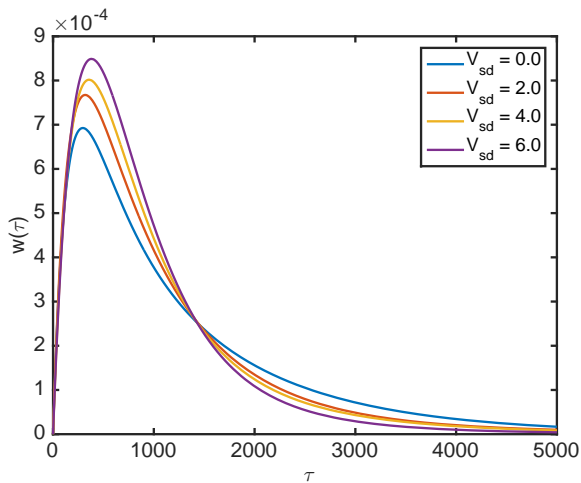}
\end{center}
	\caption{WTD between the detection of the transferred electron in the drain electrode $w_-^D(\tau)$ in strong electron-phonon coupling regime. Parameters used in calculations (all energy values are given in units of $\omega$): $\gamma_S = \gamma_D =0.01$, $ T=0.05$, 
$\epsilon =0$, $\lambda=3 $. Time $\tau$ is measured in  $1/\omega$.}	
\label{wtd2-fig}
\end{figure}

\begin{figure}[t!]
\begin{center}
\includegraphics[width=\columnwidth]{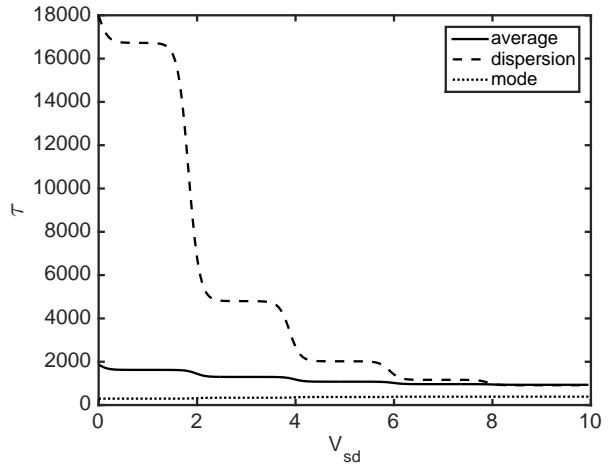}
\end{center}

\caption{Average waiting time, dispersion and  mode time  between the detection of the transferred electron in the drain electrode. Parameters used in calculations (all energy values are given in units of $\omega$): $\gamma_S = \gamma_D =0.01$, $ T=0.05$, 
$\epsilon =0$, $\lambda=3 $. Time $\tau$ is measured in  $1/\omega$ and  values of voltage bias $V_{sd}$ are given in $\omega$.}
	
\label{avt2}
\end{figure}

The increase of electron-vibration interaction  from moderate   to strong coupling regime leads to some interesting changes in the waiting time behaviour.  Fig.\ref{wtd2-fig} shows the WTD for electron tunnelling from the the molecule to the drain electrode $w_-^D(\tau)$ computed at  $\lambda/\omega =3 $. The values of the probability distribution at the peak is reduced  and the distribution is more shifted towards the larger waiting times, indicating that it takes longer for electron to transverse the molecule when the electron is strongly coupled to the molecular vibration.
As seen in Fig. \ref{avt2} the mode of the distribution shows almost no voltage dependence and the average waiting time demonstrates step-wise decrease similar to the  moderate electron-vibration interaction $\lambda/\omega =1$ case. The main changes in WTD is the dramatic reduction of the dispersion of the waiting time for tunnelling electrons when we move away from the equilibrium by increasing the applied voltage bias.

\begin{figure}[t!]
\begin{center}
\includegraphics[width=\columnwidth]{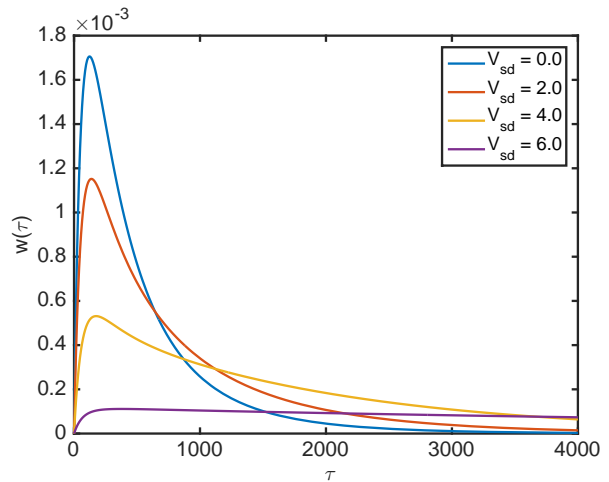}
\end{center}
	\caption{WTD between the electron transport against the average current flow, from drain electrode to the molecule,  $w_+^D(\tau)$. Parameters used in calculations (all energy values are given in units of $\omega$): $\gamma_S = \gamma_D =0.01$, $ T_S = T_D =0.05$, 
$\epsilon =0$, $\lambda=1 $. Time $\tau$ is measured in  $1/\omega$.}	
\label{wtd3-fig}
\end{figure}

Fig.\ref{wtd3-fig} shows the WTD $w_+^D(\tau)$. It describes statistics of  extreme events when  electrons tunnel against the applied voltage bias from the drain electrode into the molecule. In equilibrium, when the voltage is zero,  $w_+^{D}$ is exactly the same as  the reverse tunnnelling process $w_-^{D}$ - grand canonical ensemble equilibrium is maintained by balancing in and out particle jumps. As the voltage increases and the molecule departs from the equilibrium, the back-tunnelling events becomes rarer and rarer, the distribution decreased and becomes more and more skewed towards the long waiting time. At the large voltage the back-tunnelling events  are completely suppressed and the average waiting time becomes infinitely large for this process.

\section{Conclusions}

We have studied WTD for electron transport through a molecular junction. 
The molecule is modelled by one molecular orbital coupled with a single localised vibration.
 We treat electron-vibration interaction exactly and the influence of molecular-electrode coupling is considered perturbatively within Born-Markov approximation. The obtained master equation is used to define 4 quantum jump operators associated with different electron tunnelling processes between the molecule  and  electrodes. We compute WTDs for these jumps operators and  study these WTDs for different strengths of electron-vibration interaction and voltages.
 
  We main observations are summarised below:
\begin{itemize}
\item
The value of waiting time between electron transfer events that appears most often in the electron transport -mode of the WTD - shows little dependence on applied voltage bias and remains approximately the same as in the equilibrium.

\item As the nonequilibrium develops (that means the increase of the voltage bias), the average value of the waiting times becomes smaller and moves closer to the mode time of the distribution. That means the skewness of the distribution is decreased with the growth of the electric current. The average waiting time shows stepwise dependence on the applied voltage. These steps are associated with the excitations of vibrational states by tunneling electrons. 

\item 
The dispersion of the  WTD drops  stepwise as a function of the increasing voltage bias. Likewise to the average time, these steps are associated with the excitations of vibrational quanta by electric current.  In the strong electron-vibration coupling regime, the abrupt  changes of the dispersion dominates the other variations in the WTD behaviour when the system departs away from the equilibrium.

\item The system develops nonequilibrium and increases electric current by reducing the "diversity" of tunnelling times for current carrying electrons - the distribution of waiting times between electron tunnelling  becomes less dispersive and less skewed.

\end{itemize}

I would like to thank Samuel Rudge for many valuable discussions

\end{document}